\documentclass 
[journal]{IEEEtran}

\IEEEoverridecommandlockouts
\usepackage{ifpdf}
\usepackage{cite}
\ifCLASSINFOpdf
   \usepackage[pdftex]{graphicx}
 \DeclareGraphicsExtensions{.eps}
\else
   \usepackage[dvips]{graphicx}
  \DeclareGraphicsExtensions{.eps}
\fi
\usepackage[cmex10]{amsmath}
\usepackage{algorithmic}
\usepackage{array}
\usepackage{mdwmath}
\usepackage{mdwtab}
\usepackage{eqparbox}
\usepackage{algorithm}
\usepackage{algorithmic}

\usepackage{url}
\usepackage{mdwmath}
\usepackage{mdwtab}
\usepackage{amsmath}

\usepackage{amssymb}
\usepackage{dsfont}
\begin{document}
\title{Adaptive Modulation in Multi-user Cognitive Radio Networks over Fading Channels}
\author{\IEEEauthorblockN{F.~Foukalas$^{*}$, T.~Khattab$^{*}$, H.Vincent~Poor$^\dag$} \\
\IEEEauthorblockA{$^*$Electrical Engineering, Qatar University, Doha, Qatar\\
Email: foukalas@qu.edu.qa, tkhattab@qu.edu.qa \\
$^\dag$Electrical Engineering, Princeton University, Princeton NJ, USA\\
Email: poor@princeton.edu}}

\maketitle

\begin{abstract}
In this paper, the performance of adaptive modulation in multi-user cognitive radio networks over fading channels is analyzed. Multi-user diversity is considered for opportunistic user selection among multiple secondary users. The analysis is obtained for Nakagami-$m$ fading channels. Both adaptive continuous rate and adaptive discrete rate schemes are analysed in opportunistic spectrum access and spectrum sharing. Numerical results are obtained and depicted to quantify the effects of multi-user fading environments on adaptive modulation operating in cognitive radio networks. 
\end{abstract}


\IEEEpeerreviewmaketitle
\section{Introduction}
Adaptive modulation has been successfully deployed in wireless communication systems providing link adaptation \cite{c1}. Using adaptive modulation, the transmission rate is adapted based on the channel conditions, which are estimated at the receiver's side and made available at the transmitter through a feedback channel. When adaptive modulation is implemented in conjunction with power control at the physical layer, a variable rate variable power (VRVP) modulation is considered \cite{c2}. Two alternative schemes of VRVP have been proposed in the literature, known as continuous rate and discrete rate. The latter is more practical from an implementation point of view.

Cognitive radio (CR) has been recently proposed for enhancing spectrum utilization of licensed wireless systems when certain conditions apply \cite{c3}. The knowledge of the channel state is very important for both types of CR networks (CRNs), known as opportunistic spectrum access (OSA) and spectrum sharing (SS) \cite{c4}. Hence, the incorporation of adaptive modulation in CRNs is possible. Recently, a few investigations of the  performance of adaptive modulation in CRNs have been accomplished. More specifically, ~\cite{c5} and ~\cite{c6} investigate adaptive modulation in SS CRNs, while ~\cite{c7} and ~\cite{c8} present a performance analysis of adaptive modulation in OSA CRNs. However, none of these works have been assumed a multi-user CRN in fading channels. \let\thefootnote\relax\footnote{This research work is supported by the Qatar National Research Fund (QNRF) under National Research Priorities Program (NPRP) Grant NPRP 09-1168-2-455.} 

In this paper, we analyze and evaluate the performance of adaptive modulation in multi-user cognitive fading environment. In particular, we analyze the spectral efficiency of CRNs that employ continuous and discrete rate types of adaptive modulation operating over Nakagami-$m$ channels assuming additionally multiple secondary users (SUs). We assume multi-user diversity (MUD) using opportunistic selection of the SU with the best signal-to-noise-ratio (SNR). Finally, we provide and discuss the results of our analysis. 

The rest of this paper is organized as follows. Section II describes the multi-user cognitive radio network model. Section III provides the performance analysis of adaptive modulation operating over multi-user cognitive radio fading channels. In Section IV, we present and discuss the obtained numerical results and in Section V we provide summary of this work.

\section{System Model} \label{system}

We assume a cognitive radio network with one secondary user transmitter (SU-Tx) and multiple secondary user receivers (SU-Rxs) denoted with $i\in {1,...,L}$ where each user $i$ is served through an opportunistic or spectrum sharing access strategy \cite{c3}. We assume that the primary network (PN) consists of one primary user transmitter (PU-Tx) and one primary user receiver (PU-Rx). Fading channels are assumed for all links. The channel gain between the SU-Tx and the $i-th$ SU-Rx is denoted as $g_{s,i}$ and its additive white Gaussian noise (AWGN) denoted as $n_{s,i}$. The average transmit power over the fading channel is $\bar{P}$, the AWGN has power density $N_0/2$ and the received bandwidth is $B$. An SU-Rx can have access to a channel if and only if a predefined maximum level on the instantaneous transmit power $P$ is achieved. This level is determined from the channel state information (CSI) which represents the minimum received SNR, $\gamma_{s,i}$ that is equal to $g_{s,i}\bar{P}/N_0B$ for a channel gain $g_{s,i}$ and a unit of bandwidth $B$. 

The transmit power $P$ is controlled based on SNR $\gamma$ using power control, and thereby we denote it as $P(\gamma)$ ~\cite{c9}. The SU-Tx uses an MUD selection strategy to select transmission to the SU-R with the best received SNR ~\cite{c10}. The channel estimate, i.e. $\gamma_{s,i}$ is also available at the SU-Tx side via a feedback channel. We assume that the CSI is perfectly available at the receivers i.e. PU-Rx and SU-Rxs, and that the feedback channel does not induce any delays on the CSI’s transmission. Moreover, a set of $M-ary$ Quadrature Amplitude Modulations (M-QAMs) is considered and their selection relies on the estimated CSI. In the considered system model, first the SU-Tx determines the user who can access the channel through the MUD and in the sequel it selects the transmission rate $R = log_2(M)$ via the selection of the appropriate $M-ary$ modulation from the signal set according to the estimated CSI. 

Finally, we make the following assumptions for the considered system: a) the transmission of each symbol is accomplished with a symbol period $T_s = 1/B$ using ideal raised cosine pulses; and b) the fading channel is varying slowly in time, i.e. the receiver is able to sense and track the channel fluctuations and thus it corresponds to a block flat fading channel model with an average received SNR, $\bar{\gamma}$ ~\cite{c11}.

\section{Performance Analysis} \label{analysis}

Assuming VPVR adaptive modulation since power control is used in both SS and OSA CRNs, we will analyse both continuous and discrete rate cases denoted as CR and DR respectively. We derive below first the channel capacity achieved over fading channels and second the spectral efficiency assuming CR and DR adaptive modulation schemes. As mentioned above, the SU-Tx employs MUD to select the SU-Rx and therefore, the received SNR of the selected SU-Rx $\gamma_{s,max}$ is obtained as follows ~\cite{c10}: 

\begin{eqnarray} \label{eq1}
 \gamma_{s,max} = \max_{1\leq i \leq L} \ \gamma_{s,i}
\end{eqnarray}
with probability density function (PDF) obtained as follows:
\begin{eqnarray} \label{eq2}
 f_{\gamma_{s,max}}(x) = L f_{\gamma_{s,i}}(x)F_{\gamma_{s,i}}(x)^{L-1}
\end{eqnarray}
where $f_{\gamma_{s,i}}(x)$ and $F_{\gamma_{s,i}}(x)$ are the PDF and the cumulative distribution function (CDF) of the received SNR $\gamma_{s,i}$ at the $i-th$ SU-Rx respectively.  
The overall average achievable capacity at the secondary system (i.e. SU-Tx to SU-Rx) is obtained as follows:
\begin{eqnarray} \label{eq3}
 C_{s} &=& \int_0^{\mathcal{\infty}} { B log_2(1+\gamma_{s,max})f_{\gamma_{s,max}}(x) dx} .
\end{eqnarray} 

\subsection{Channel Capacity} \label{CR}

The average channel capacity of a fading channel $\bar{C}$ (in bits per second) is given by ~\cite{c9} 

\begin{eqnarray} \label{eq4}
\bar{C} = \max_{P(\gamma)} \left\lbrace \int_{0}^{\infty}  B log_2 \left( 1+ \gamma \frac{P(\gamma)}{\bar{P}} \right) f(\gamma) \,d \gamma \right\rbrace 
\end{eqnarray}
where the instantaneous transmit power $P(\gamma)$ chosen relative to $\gamma$ is subject to the following power constraints:
\begin{eqnarray} \label{eq5}
\int_{0}^{\infty}  P(\gamma) f(\gamma) \,d \gamma \leq \bar{P} 
\end{eqnarray}

\begin{eqnarray} \label{eq6}
\int_{0}^{\infty}  P(\gamma_{sp}) f(\gamma_{sp}) \,d \gamma_{sp} \leq \bar{Q} 
\end{eqnarray}
where \eqref{eq5} represents the well-known transmit power constraint applied to OSA systems and the \eqref{eq6} represents the additional interference power constraint applied to SS systems ~\cite{c4}.  

\subsubsection{Transmit Power Constraint} \label{constr1}

We consider the case of the average transmit power constraint, in which the fading distribution depends only on secondary link and  the optimal power allocation of the SU-Tx is obtained as follows \cite{c4}:
\begin{eqnarray} \label{eq7}
\frac{P(\gamma_s)}{\bar{P}}  = \left[ \frac{1}{\gamma_{0,s}} - \frac{1}{\gamma_s} \right] ,& if & \gamma_s > \gamma_{0,s} 
\end{eqnarray}
where $\gamma_{0,s}$ is the optimal cut-off level of the received SNR at the SU-Rx, which can be calculated by the substitution of \eqref{eq7} into \eqref{eq5} with equality for maximizing the capacity in \eqref{eq4}. 

Considering MUD in conjunction with the average transmit power constraint, the capacity is obtained as follows:
\begin{eqnarray} \label{eq8}
\bar{C} = \int_{\gamma_{0,s}}^{\infty}  B log_2(\frac{\gamma_{s,max}}{\gamma_{0,s}}) f(\gamma_{s,max}) \,d \gamma_{s,max} .
\end{eqnarray}

\subsubsection{Interference Power Constraint} \label{constr2}

We consider now the case of the average interference power constraint, in which the fading distribution depends on both secondary and interference links and the optimal power allocation of the SU-Tx is obtained as follows \cite{c4}:
\begin{eqnarray} \label{eq9}
\frac{P(\gamma_{sp})}{\bar{P}}  = \left[ \frac{1}{\gamma_{0,sp}} - \frac{1}{\gamma_{sp}} \right], & if & \gamma_{sp} > \gamma_{0,sp} 
\end{eqnarray}
where $\gamma_{0,sp}$ is the optimal cut-off level of the received SNR at the SU-Rx considering the interference power constraint and thereby the $\gamma_{sp}$ is equal to \cite{c4}:
\begin{eqnarray} \label{eq10}
\gamma_{sp} = \frac{g_{s,i}\bar{P}}{g_p N_0 B} .
\end{eqnarray}
Considering again MUD in conjunction with the average transmit power constraint, the capacity is obtained as follows:
\begin{eqnarray} \label{eq11}
\bar{C} = \int_{\gamma_{0,s}}^{\infty}  B log_2(\frac{\gamma_{sp,max}}{\gamma_{0,sp}}) f(\gamma_{sp,max}) \,d \gamma_{sp,max} .
\end{eqnarray}
Notably, the PDFs $f(\gamma_{s,max})$ and $f(\gamma_{sp,max})$ will be obtained for Rayleigh and Nakagami-$m$ distribution using the analysis provided in Section IV whereby the PDF and CDF will be obtained for a single user and then using equation \eqref{eq2} for the one with the best SNR. 

\subsection{Spectral Efficiency in Continuous Rate Adaptive Modulation} \label{Se}

\subsubsection{Transmit Power Constraint}

The power allocation that maximizes the spectral efficiency in SS system, i.e. assuming \eqref{eq7} and the adaptive modulation in ~\cite{c2}, is given as follows: 
\begin{eqnarray}\label{eq12}
\frac{P(\gamma_s)}{\bar{P}}=
\begin{cases}
\ \frac{1}{\gamma_{0,s}}-\frac{1}{\gamma_s K}, \gamma_s \geq \frac{\gamma_{0,s}}{K} \\
\ 0 , \gamma_s < \frac{\gamma_{0,s}}{K}\\ 
\end{cases}
\end{eqnarray}
where $K$ is an effective power loss that retains the bit-error-rate (BER) value and is equal to:
\begin{eqnarray} \label{eq13}
 K = \frac{-1.5}{ln(5BER)} .
\end{eqnarray}
Combining the equations above, the spectral efficiency for the continuous rate adaptive modulation is maximized up to a cut-off level in SNR denoted as $\gamma_{0,s,K}=\gamma_s/K$ obtained as follows \cite{c2}:
\begin{eqnarray} \label{eq14}
 \langle S_e \rangle_{CR} = \int_{\gamma_{s,K}}^{\infty} log_2(\frac{\gamma_{s,max}}{\gamma_{0,s,K}})f(\gamma_{s,max})d\gamma_{s,max} .
\end{eqnarray}

\subsubsection{Interference Power Constraint}

In the same way as above and taking into account \eqref{eq9}, we have the following:
\begin{eqnarray}\label{eq15}
\frac{P(\gamma_{sp})}{\bar{P}}=
\begin{cases}
\ \frac{1}{\gamma_{0,sp}}-\frac{1}{\gamma_{sp} K}, \gamma_{sp} \geq \frac{\gamma_{0,sp}}{K} \\ 
\ 0 , \gamma_{sp} < \frac{\gamma_{0,sp}}{K} . \\ 
\end{cases}
\end{eqnarray}
Replacing the index $s$ with the index $sp$ in \eqref{eq14} taking into account \eqref{eq15}, we can find $S_e$ at the link with channel gain $g_{sp}$. Again, the $f(\gamma_{sp,max})$ is obtained using \eqref{eq2} and the analysis obtained in Section IV.
 
\subsection{Spectral Efficiency in Discrete Rate Adaptive Modulation}

We now consider a DR MQAM with a constellation set of size $N$ with $M_0 = 0$,$M_1 = 2$ and $M_j = 2^{2(j-1)}$ for $j = 2,...,N$. At each symbol time, the system transmits with a constellation from the set ${M_j = 0,1,...,N}$ ~\cite{c2}. The choice of a constellation depends on $\gamma$, i.e. the SNR over that symbol time, while the $M_0$ constellation corresponds to no data transmission. The spectral efficiency is now defined as the sum of the data rates of each constellation multiplied by the probability that this constellation will be selected and thus it is given as follows:
\begin{eqnarray} \label{eq16}
 \langle Se \rangle_{DR} = \Sigma_{j=1}^{N}log_2(M_j)f(\gamma_{s,j} \leq \gamma \leq \gamma_{s,j+1})
\end{eqnarray}
subject to the following power constraint:
\begin{eqnarray} \label{eq17}
 \Sigma_{j=1}^{N} \int_{\gamma_{s,j}}^{\gamma_{s,j+1}} \frac{P_j(\gamma)}{\bar{P}} p(\gamma_s)d\gamma_s = 1
\end{eqnarray}
where $P_j(\gamma_s)/\bar{P}$ is the optimal power allocation that is obtained from (\ref{eq7}) for each constellation $M_j$ with a fixed BER as follows:
\begin{eqnarray}\label{eq18}
\frac{P_j(\gamma_s)}{\bar{P}}=
\begin{cases}
\ (M_j-1)\frac{1}{\gamma_{s,K}} -\frac{1}{\gamma_s K}, M_j \leq \frac{\gamma_s}{\gamma_{s,K}^*} \leq M_{j+1}\\
\ 0 , M_j=0\\
\end{cases}
\end{eqnarray}
where $\gamma_{s,K}^*$ is a parameter that will later be optimized to maximize spectral efficiency by defining the optimal constellation size for each $\gamma_s$. The analysis for the interference power constraint is obtained as above by replacing $\gamma_s$ with $\gamma_{sp}$ taking into account (9) and (10). 

%
%
%
%
%

\section{Fading Distributions in Multi-User Environments} \label{Fadings}
\subsection{Rayleigh Distribution} \label{Rayl2} 
\subsubsection{Opportunistic Spectrum Access}

We assume that the channel gains $g_{s,i}$ and $g_{p}$ are independent and identically distributed (i.i.d.) Rayleigh random variables  $\forall i$. In OSA systems, only the transmit power constraint is applied and thus \eqref{eq7} depends on the channel gain on the secondary links i.e. $g_{s,i}$; and the PDF is obtained as follows \cite{c9}:
\begin{eqnarray} \label{eq22} 
f(x) &=& e^{-x} .
\end{eqnarray}
and the CDF of the PDF in \eqref{eq22} is obtained as follows: 
\begin{eqnarray} \label{eq23} 
F(x) = \frac{1}{log(e)} \left( 1-e^{-x} \right) .
\end{eqnarray}

Substituting \eqref{eq22} and \eqref{eq23} into \eqref{eq2}, we can derive the PDF $f_{\gamma_{s,max}}(x)$ of the maximum received SNR and thus the capacity and spectral efficiency for CR and DR adaptive modulations derived above. 

\subsubsection{Spectrum Sharing}

We assume that the channel gains $g_{s,i}$ and $g_{p}$ are i.i.d. Rayleigh random variables  $\forall i$. For notational brevity, we will denote the term $g_{s,max}/g_{p}$ as $g_s/g_p$. We will substitute $X=g_s/g_p$ so that the PDF of the received SNR at the SU-Tx is obtained as follows: 

\begin{eqnarray} \label{eq24} 
\nonumber
f(x) &=& \int_{0}^{\infty} \ z e^{-x z} e^{-z}  \,d z\\ 
&=& -\frac{e^{-(1+x)z}(1+z+x z)}{(1+x)^2}\mid_0^\infty = \frac{1}{(1+x)^2}
\end{eqnarray}
which is identical to the expression presented in \cite{c14}. The CDF of the PDF in \eqref{eq24} is obtained as follows: 

\begin{eqnarray} \label{eq25} 
F(x) = 1 - \frac{1}{1+x} . 
\end{eqnarray}
Substituting \eqref{eq24} and \eqref{eq25} into \eqref{eq2}, we can derive the PDF $f_{\gamma_{s,max}}(x)$ of the maximum received SNR and thus the capacity and spectral efficiency for CR and DR adaptive modulations derived above. 

\subsection{Nakagami$-m$ Distribution} \label{Nakag} 
\subsubsection{Opportunistic Spectrum Access}
We now assume that the channels gains $g_{s,i}$ and $g_{p}$ are i.i.d. Nakagami$-m$ random variables $\forall i$ and thus the following Nakagami$-m$ distribution is applied: 

\begin{eqnarray} \label{eq26}
f(x) = \frac{m^m x^{(m-1)}}{\Gamma(m)}e^{-m )}, & &  x\geq 0
\end{eqnarray}
and the CDF of the PDF in \eqref{eq26} is obtained as follows: 
\begin{eqnarray} \label{eq27} 
F(x) = \frac{(m-1)m^{m-2}}{\Gamma(m)} (1-(1+m x)e^{-m x}) .
\end{eqnarray}

\subsubsection{Spectrum Sharing}

We now assume that the channels gains $g_{s,i}$ and $g_{p}$ are i.i.d. Nakagami$-m$ random variables $\forall i$ and thus follow the following Nakagami$-m$ distribution for a specific channel gain $Z=z$ : 
\begin{eqnarray} \label{eq28}
f(z) = \frac{m^m z^{(m-1)}}{\Gamma(m)}e^{(-m z)}, & &  z\geq 0
\end{eqnarray}
where $m$ represents the shape factor under which the ratio of the line-of-sight (LoS) to the multi-path component is realized \cite{c15}. Assuming that both channel gains $g_{s,i}$ and $g_{p}$ have instantaneously the same fading fluctuations i.e. $m_s=m_p=m$, the PDF of the term $X=g_s/g_p$ is obtained as follows: 

\begin{eqnarray} \label{eq29} 
f(x) = \frac{x^{m-1}}{B(m,m)(x+1)^{2m}}, & & x \geq 0 . 
\end{eqnarray}

After some mathematical manipulation, the CDF of the PDF in \eqref{eq29} is obtained as follows: 

\begin{eqnarray} \label{eq30} 
F_{g_s/g_p}(x) = \frac{1}{B(m,m)} \frac{x^m}{m} {_2}F_1(m,2m;1+m;-x)
\end{eqnarray}
where ${_2}F_1(a,b;c;y)$ is the Gauss hyper-geometric function which is a special function of the hyper-geometric series \cite{c16}. Substituting \eqref{eq29} and \eqref{eq30} into \eqref{eq2}, we can derive the PDF of the received SNR $\gamma_{s,max}$ of the selected SU-Rx for the Nakagami-$m$ case.  

\section{Numerical Results}

In the following figures, we depict the capacity and spectral efficiency for continuous and discrete rate cases respectively in OSA and SS CRNs. More specifically, Fig.1 depicts the capacity and spectral efficiency for continuous and discrete rate in $bits/Hz$ versus the average transmit power $P_{av}$ in the secondary link for different number of secondary users (i.e. SU-Rxs) equal to $N_s=1$, $N_s=5$ and $N_s=15$. In this figure, the average interference power constraint does not exist, in other words, we use the transmit power constraint applied in OSA CRNs. Thereby, the performance of adaptive modulation in OSA CRNs is depicted with multiple SU-Rxs. For the discrete rate case, we assume 5 regions of M-QAM with $M\in[{0,4,8,16,64}]$. Obviously, as long as the number of secondary users $N_s$ increases, the capacity and spectral efficiency increase as well. We notice that a small increase in the number of SU-Rxs i.e. $N_s=5$ gives a big performance enhancement in $bits/Hz$, almost more than one and half times. However, a bigger increase in SU-Rxs i.e. $N_s=15$ gives smaller performance enhancement, indicating thereby that the increase in capacity and spectral efficiency exhibits a saturation behavior. 

Fig.2 depicts the capacity and spectral efficiency over the average interference power $Q_{av}$ at the link between the SU-Tx and PU-Rx. The average transmit power is taken to be $P_{av}=20dB$. Thereby, the performance of adaptive modulation in SS CRNs is depicted for multiple SU-Rxs over the interference channel. We realize that the performance increase is higher than the one over the average transmit power $P_{av}$ as depicted in Fig.1 either assuming $N_s=5$ or $N_s=15$. This is due to the fact that as the number of SU-Rxs increases, the possibility of finding an SU-Rx with sufficient SNR increases and thus the performance rate on the interference link increases as long as the constraint is being relaxed, i.e. $Q_{av}$ increases. 
\begin{figure}[ht]
		\includegraphics[width=9cm,height=7cm]{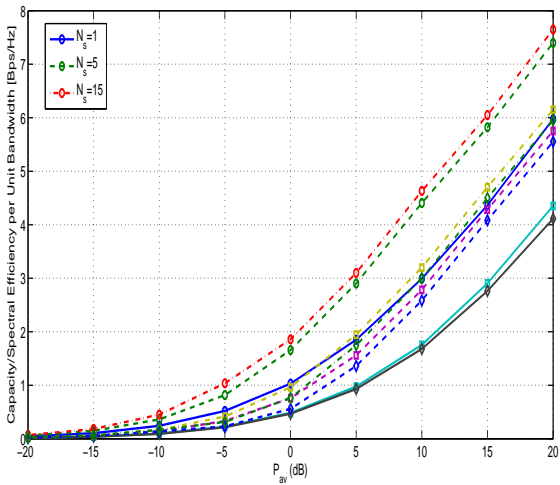}
	\caption{Channel capacity ($-o$) and spectral efficiency of continuous ($-s$) and discrete rate ($-d$) adaptive modulation vs. the average transmit power $P_{av}$ for different number of secondary users $N_s$ as depicted.} 
	\label{fig:model-system}
\end{figure}
\begin{figure}[ht]
		\includegraphics[width=9cm,height=7cm]{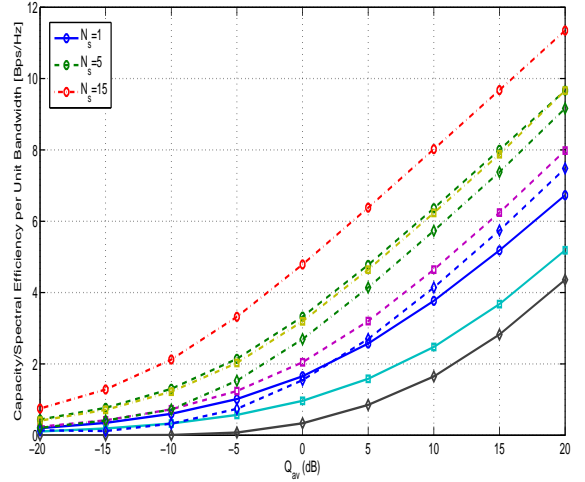}
	\caption{Channel capacity ($-o$) and spectral efficiency of continuous ($-s$) and discrete rate ($-d$) adaptive modulation vs. the average interference power $Q_{av}$ for different number of secondary users $N_s$ as depicted.} 
	\label{fig:model-system}
\end{figure}

Fig. 3 depicts the spectral efficiency in continuous and discrete rate versus the number of secondary users (SU-Rxs) with the Nakagami-$m$ fading coefficient given by m = 1 assuming Rayleigh and m = 2 assuming Ricean factor equal to 2.4312. In addition we assume  interference power constraints of $Q_{av}=-10dB$, $Q_{av}=0dB$ and $Q_{av}=10dB$ as well as a transmit power of $P_{av}=10dB$. Thereby, we depict the performance of adaptive modulation in SS CRNs versus the number of secondary users i.e. SU-Rxs. The impact of $m$ is more evident for high interference power constraints e.g. $Q_{av}=10dB$, where the degradation from $m=1$ to $m=2$ can be more than $2Bps/Hz$ for high number of SU-Rxs, e.g. $N_s=15$. On the other hand, the impact is negligible for low average interference power constraints, e.g. $Q_{av}=-10dB$, where the fading environment i.e. changes in $m$  does not decrease the performance significantly. For a more comprehensive view in Nakagami-$m$ channels, we depict in Fig.4 the spectral efficiency vs. the average interference power $Q_{av}$ for average transmit power $P_{av}=10dB$, and thereby the case of a SS CRN, different number of secondary users $N_s=5$ and $N_s=15$ for $m = 1$ (Rayleigh) and $m = 2$ (Ricean) for the Nakagami−$m$ distribution. Obviously, in Rayleigh conditions the system achieves better performance and the gain is more evident when the number of secondary users $N_s$ increases. 

\begin{figure}[ht]
		\includegraphics[width=9cm,height=8cm]{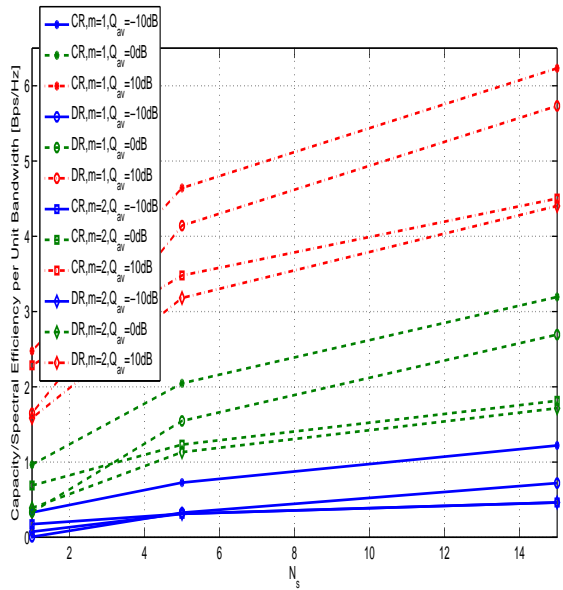}
	\caption{Capacity and spectral efficiency vs. number of SU-Rxs $N_s$ with $m = 1$ (Rayleigh) and $m = 2$ (Ricean) for the Nakagami−m distribution.}
	\label{fig:model-system}
\end{figure}
\begin{figure}[ht]
		\includegraphics[width=9cm,height=8cm]{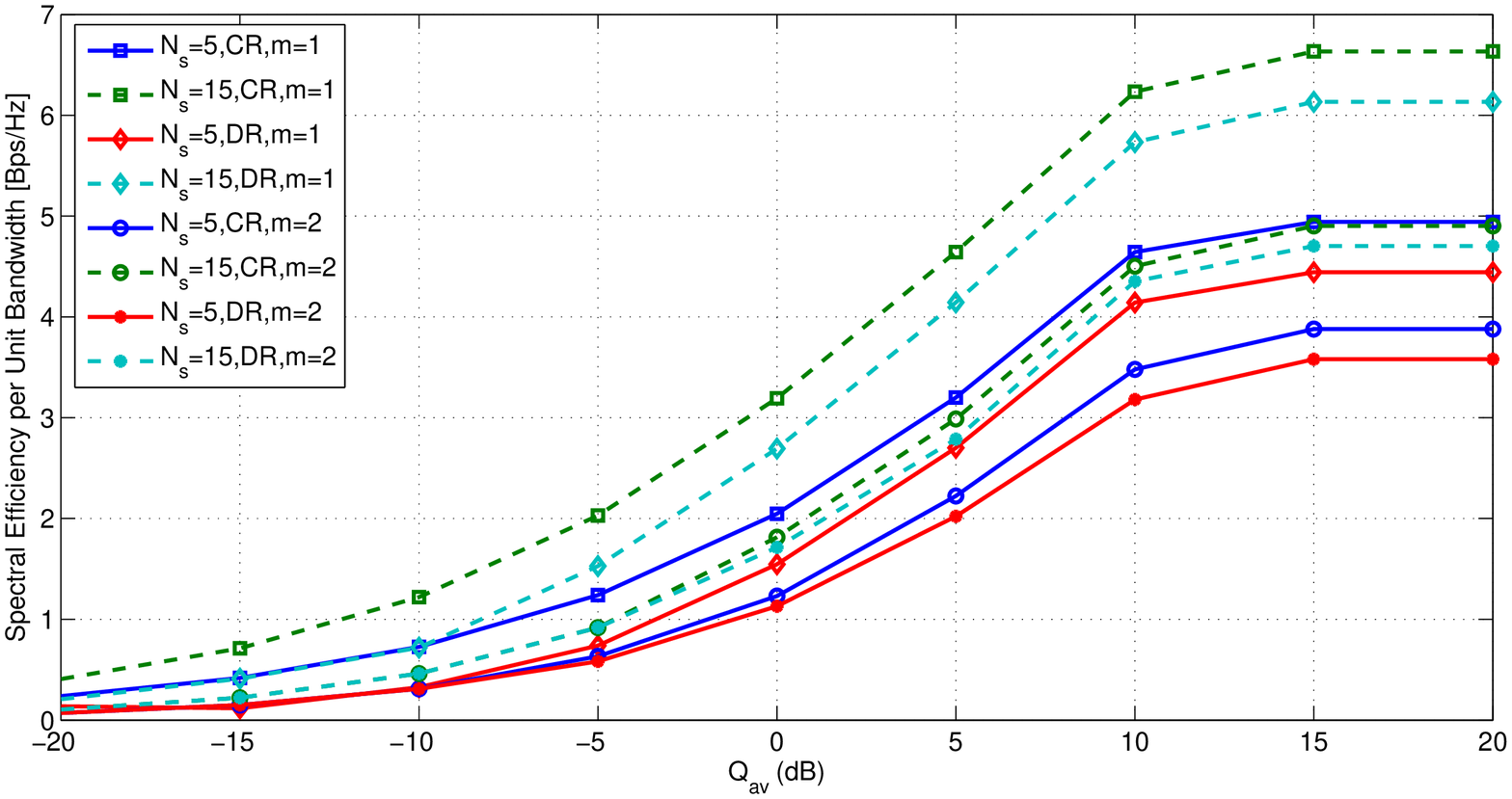}
	\caption{Spectral efficiency vs. the average interference power $Q_{av}$ for average transmit power $P_{av}=10dB$, different number of secondary users $N_s$ and with $m = 1$ (Rayleigh) and $m = 2$ (Ricean) for the Nakagami−m distribution.}
	\label{fig:model-system}
\end{figure}

\section{Summary}
In this work, we have analyzed adaptive modulation in multi-user cognitive radio fading environments. In particular, we have analyzed the performance of adaptive modulation in cognitive radio networks with multiple secondary users assuming multi-user diversity as a transmission selection strategy. Both opportunistic spectrum access and spectrum sharing cognitive radio systems are considered using constraints on the transmit and interference power, respectively. The derived fading distributions model both Rayleigh and Nakagami-$m$ channels. Finally, the spectral efficiency gain is depicted in a multiple secondary user environment.

\end{document}